\documentclass[final,5p,times,twocolumn]{elsarticle}
\usepackage[english]{babel}
\usepackage{amsmath,amssymb,graphicx,hyperref}
\usepackage{xcolor}
\usepackage[utf8]{inputenc}
\usepackage{listings}
\usepackage{xcolor}
\usepackage{hyperref}
\hypersetup{colorlinks=true, citecolor=blue, filecolor=blue, linkcolor=blue, urlcolor=blue}
\newcommand{\cdummy}{\cdot}
\newcommand{\nocomma}{}
\newcommand{\tmcolor}[2]{{\color{#1}{#2}}}
\newcommand{\tmem}[1]{{\em #1\/}}
\newcommand{\tmmathbf}[1]{\ensuremath{\boldsymbol{#1}}}
\newcommand{\tmop}[1]{\ensuremath{\operatorname{#1}}}

\newcommand{\tmsamp}[1]{\texttt{#1}}

\newcommand{\tmverbatim}[1]{{\ttfamily{#1}}}
\definecolor{pastelyellow}{HTML}{FFFFDF}
\journal{Computer Physics Communications}

\begin{document}

\begin{frontmatter}

\title{\textsc{ElasTool}: An automated toolkit for elastic constants calculation}

\author{Zhong-Li Liu}

\cortext[author]{\textit{E-mail address:} zl.liu@163.com}
\address{College of Physics and Electric Information, Luoyang Normal
	University, Luoyang 471934, China}

\begin{abstract}
We here present, the \textsc{ElasTool} package, an automated toolkit for
calculating the second-order elastic constants (SOECs) of any two- (2D) and three-dimensional (3D) crystal
systems. It can utilize three kinds
of strain-matrix sets, the high-efficiency strain-matrix sets (OHESS), the
universal linear-independent coupling strains (ULICS) and the all-single-element strain-matrix sets (ASESS) to calculate the SOECs
automatically. In an automatic manner, \textsc{ElasTool} can deal with
both zero- and high-temperature elastic constants. The theoretical
background and computational method of elastic constants, the package
structure, the installation and run, the input/output files, the controlling
parameters, and two representative examples of \textsc{ElasTool} are
described detailedly. \textsc{ElasTool} is useful for either the exploration of
materials' elastic properties or high-throughput new materials design.
\textsc{ElasTool} is also available at our website:
\href{http://www.matdesign.cn/}{www.matdesign.cn}.

\end{abstract}


\end{frontmatter}

%\linenumbers
\noindent
{\bf PROGRAM SUMMARY}\\
\begin{small}
\noindent
{\em Program Title:} ElasTool \\
{\em Licensing provisions:} GNU General Public License, version 3\\
{\em Programming language:} Python 3 \\
{\em Computer:} Any computer that can run Python (versions 3.5 and later).\\
{\em Operating system:} Any operating system that can run Python.\\
{\em External routines:} NumPy \cite{1}, Spglib \cite{2}, ASE \cite{3}, Pandas \cite{4}\\
{\em Nature of problem:} The stress-strain method of elastic constants calculation
depends on more accurate stresses calculated by density functional theory (DFT) compared to the strain-energy method. But its advantage is that it needs a smaller number of strain sets to solve the equation sets to deduce elastic constants and more straightforward to implement. While the more accurate stresses take much more time in DFT calculating. Thus, a smaller number of strain sets and more efficient strain sets are urgently needed to improve the computational efficiency of elastic constants. An automated solution coupled with DFT is necessary for the exploration of materials' elastic properties and high-throughput new materials design.\\
{\em Solution method:} The solution to improving the computational efficiency of the stress-strain method is to decrease the number of strain-matrix sets and optimize the strain-matrix sets.  We coupled our previously proposed high-efficiency strain-matrix sets (OHESS) with DFT and automated the processes of the calculation of the elastic constants using the stress-strain method in the \textsc{ElasTool} package. \textsc{ElasTool} can also adopt the all-single-element strain-matrix sets (ASESS) and the universal linear-independent coupling strains (ULICS). It can deal with both zero- and high-temperature elastic constants of any crystal systems belonging to 2D or 3D. It can also output other elastic moduli, the elastic anisotropy, Debye temperature, the sound velocities, and elastic stability based on the calculated elastic constants.\\
{\em Additional comments:} This package can use the stress tensors calculated by other DFT codes, such as the Vienna Ab initio Simulation Package (VASP) \cite{5,6,7}.\\
{\em Running time:} The time used by the examples provided in the distribution mainly spends on the DFT parallel running.
\  %Provide any additional comments here.
   \\

\end{small}

\section{Introduction}
Elastic constants are fundamental parameters of materials, which are crucial
to many disciplines, including physics {\cite{Ashcroft1976}}, condensed matter,
materials science (Ref. {\cite{Tehrani2019}}), geophysics {\cite{Lay1995}},
and chemical {\cite{Tatsumi2003}}. In new materials' theoretical design aided
by crystal structure prediction {\cite{Liu2014,Wang2010,Wang2012,Glass2006,Lonie2011,Pickard2011}}, elastic
constants are frequently used to check the stability of the predicted
structures, according to Born elastic stability criteria
{\cite{Wu2007,Mouhat2014,Nye1985,Born1954,Liu2017,Yu2010}}. More
interestingly, for the informatics-based new materials screening and
design, elastic constants are important properties and have been extensively
calculated and collected in some materials database, such as the Materials
Project (MP) {\cite{Jong2015}} and JARVIS-DFT {\cite{Choudhary2018}}
databases.

The high-accuracy state of the art density functional theory (DFT)
{\cite{Hohenberg1964,Kohn1965}} is capable of calculating elastic constants
accurately. For the calculation method, generally, there are two ways to
calculate elastic constants, the energy-strain method and the stress-strain
method (Refs. {\cite{Wang2019,Golesorkhtabar2013,Perger2009}} and references
therein). Compared to the former, the latter depends on highly accurate stress
tensors but a smaller number of strain sets
{\cite{Golesorkhtabar2013,Wang2019}}. The high-accuracy stress tensors
calculation always needs higher energy cutoff and denser K-point meshes.
Although it is much more computationally expensive and time-consuming, it is
simple to implement and straightforward to calculate the high-pressure elastic
constants of materials without pressure corrections, unlike the
the strain-energy method which needs complex pressure corrections {\cite{Sinko2002}}.

Our previously proposed OHESS can keep
the symmetry of strained crystal to the most extent and largely lower
computation time at the same level of accuracy {\cite{Liu2020}}. This method
is integrated into the \textsc{ElasTool} package, which can calculate the
SOECs of any crystal systems belonging to 2D and
3D. \textsc{ElasTool} can also uses the ULICS {\cite{Yu2010}} to calculate
elastic constants. However, in the ULICS, the coupling of stress components by
including several strain components at the same time will largely reduce the
symmetry of the strained crystal. This will greatly lengthen the computation time
{\cite{Liu2020}}. \textsc{ElasTool} has also integrated the
ASESS, in which only one strain
component is applied on crystal lattice in a certain strain-matrix set.

The rest of this paper is arranged as follows. The theoretical background of
elasticity is presented in the next section. The descriptions of the package
are detailed in Sec.~\ref{desp}. The input and output files are described in
Sec.~\ref{iop}. Input parameters are explained in Sec.~\ref{ip}. The calculation
examples are shown in Sec.~\ref{exa}. Section~\ref{conclusion} is a summary of
this paper.

\section{Theoretical background of elasticity}\label{method}

\subsection{ Elastic constants computation method}

Within the linear elastic regime of a crystal, the relation between the
stresses $\sigma_i$ and corresponding strains $\varepsilon_j$ conforms to
Hooke's law,
\begin{equation}
\sigma_i = \sum_{j = 1}^6 C_{i j} \varepsilon_j \label{hooke}
\end{equation}
where the coefficient $C_{i j}$ are the elastic stiffness constants of the
crystal.

If we deform a crystal by applying the strain $\varepsilon_j$ and calculate
the corresponding stresses, we can obtain its elastic constants from
Eq.\ref{hooke}. The deformation matrix applied to the crystal unit cell is
\begin{equation}
\tmmathbf{D}=\tmmathbf{I}+\tmmathbf{\varepsilon},
\end{equation}
where $\tmmathbf{I}$ is the $3 \times 3$ unit matrix, and
$\tmmathbf{\varepsilon}$ is the strain-matrix notated by Voigt method. In the
3D case, the strain matrix is
\begin{equation}
\tmmathbf{\varepsilon}= \left[ \begin{array}{l}
\varepsilon_1 \quad {\varepsilon_6}/{2} \quad
{\varepsilon_5}/{2}\\
{\varepsilon_6}/{2} \quad \varepsilon_2 \quad
{\varepsilon_4}/{2}\\
{\varepsilon_5}/{2} \quad {\varepsilon_4}/{2} \quad \varepsilon_3
\end{array} \right] .
\end{equation}

For the 2D layered crystal, \textsc{ElasTool} assumes the crystal plane in
the $x y$ plane and the strain matrix is
\begin{equation}
\tmmathbf{\varepsilon}= \left[ \begin{array}{l}
\varepsilon_1 \quad {\varepsilon_6}/{2} \quad 0\\
{\varepsilon_6}/{2} \quad \varepsilon_2 \quad 0\\
0 \qquad 0 \quad 0
\end{array} \right] .
\end{equation}
After deformation, the crystal lattice vector is
\begin{equation}
\tmmathbf{A}' =\tmmathbf{A} \cdummy \tmmathbf{D}
\end{equation}
where $\tmmathbf{A}$ is the original crystal lattice vector.

The SOECs can be readily derived by fitting the first-order function to the
stress-strain data according to Eq.\ref{hooke}.

\subsection{Strain matrix sets}

In order to extract all SOECs of a crystal belonging to a specific crystal
system, we need deformation matrices defined by a set of strain matrices to
apply on the crystal lattice. We previously proposed the OHESS and finally
improved the computational efficiency of elastic constants considerably, as
reported in Ref.{~\cite{Liu2020}}. \textsc{ElasTool} uses three kinds of
strain-matrix sets, the OHESS, the ASESS, and the ULICS to solve the
stress-strain equation sets to find the corresponding elastic constants
numerically. The ASESS is defined in Ref.~{\cite{Liu2020}}. As for the details
of the ULICS, the readers are referred to as Ref.~{\cite{Yu2010}}. Overall, the
OHESS has the highest efficiency among the three kinds of strain-matrix sets, as
has been tested in Ref.{~\cite{Liu2020}}.

\subsection{Elastic moduli}

From the calculated elastic constants, we can derive other elastic moduli
easily. The Voigt and Reuss bulk and shear moduli for different crystal
systems are calculated according to Ref. {\cite{Wu2007}}. From the
Voigt--Reuss--Hill approximations {\cite{Hill1952}}, the arithmetic average of
Voigt and Reuss bounds is written as
\begin{equation}
B_{\tmop{VRH}} = \frac{B_V + B_R}{2}, \label{bvrh}
\end{equation}
and
\begin{equation}
G_{\tmop{VRH}} = \frac{G_V + G_R}{2},
\end{equation}
respectively.

Young's modulus $E$ is obtained by
\begin{equation}
E = \frac{9 B \nocomma G}{3 B + G},
\end{equation}
and Poisson's ratio is
\begin{equation}
\nu = \frac{3 B - 2 G}{2 (3 B + G)} . \label{nu}
\end{equation}
The elastic anisotropy is a crucial measurement of anisotropy of chemical
bonding. The Chung-Buessem anisotropy index {\cite{Chung1967}} is defined as
\begin{equation}
A^C = \frac{G_V - G_R}{G_V + G_R} \label{ac}
\end{equation}
The universal elastic anisotropy index proposed by Ranganathan and
Ostoja-Starzewski {\cite{Ranganathan2008}} is
\begin{equation}
A^U = 5 \frac{G_V}{G_R} - \frac{B_V}{C_R} - 6 \geqslant 0 \label{au}
\end{equation}

\subsection{Sound velocity}

The phase velocity $v$ and polarization of the three waves along a fixed
propagation direction defined by the unit vector $n_i$ are given by Cristoffel
equation
\begin{equation}
(C_{i \nocomma j \nocomma k \nocomma l} n_j n_k - \rho v^2 \delta_{i
	\nocomma j}) u_i = 0
\end{equation}
where $C_{i \nocomma j \nocomma k \nocomma l}$ is the
fourth-rank tensor description of the elastic constants, $n$ is the
propagation direction, and $u$ the polarization vector.

The longitudinal wave velocity is defined by
\begin{equation}
v_P = \sqrt{\frac{B + 4 G / 3}{\rho}}, \label{vp}
\end{equation}
and the body wave velocity is
\begin{equation}
v_B = \sqrt{\frac{B}{\rho}} .
\end{equation}
The shear wave velocity is written as
\begin{equation}
v_S = \sqrt{\frac{G}{\rho}} .
\end{equation}
From $v_S$ and $v_P$, we can average wave velocity via
\begin{equation}
v_m = \left[ \frac{1}{3} \left( \frac{2}{v_S^3} + \frac{1}{v_P^3} \right)
\right]^{- 1 / 3}. \label{vm}
\end{equation}
From elastic constant data we can calculate the Debye temperature from the
average sound velocity $v_m$ via
\begin{equation}
\Theta_D = \frac{h}{k} \left[ \frac{3 n}{4 \pi} \left( \frac{N_A \rho}{M}
\right) \right]^{1 / 3} v_m \label{td}
\end{equation}
where $h$ is Planck's constant, $k$ the Boltzmann's constant, $N_A$ the
Avogadro's number, $n$ the number of atoms in the unit cell, $M$ the weight of
the unit cell, and $\rho$ the density.

\section{Description of the package}\label{desp}

\subsection{The flowchart of ElasTool}

\begin{figure*}[!htbp]
	\centering
	\includegraphics[width=0.75\textwidth]{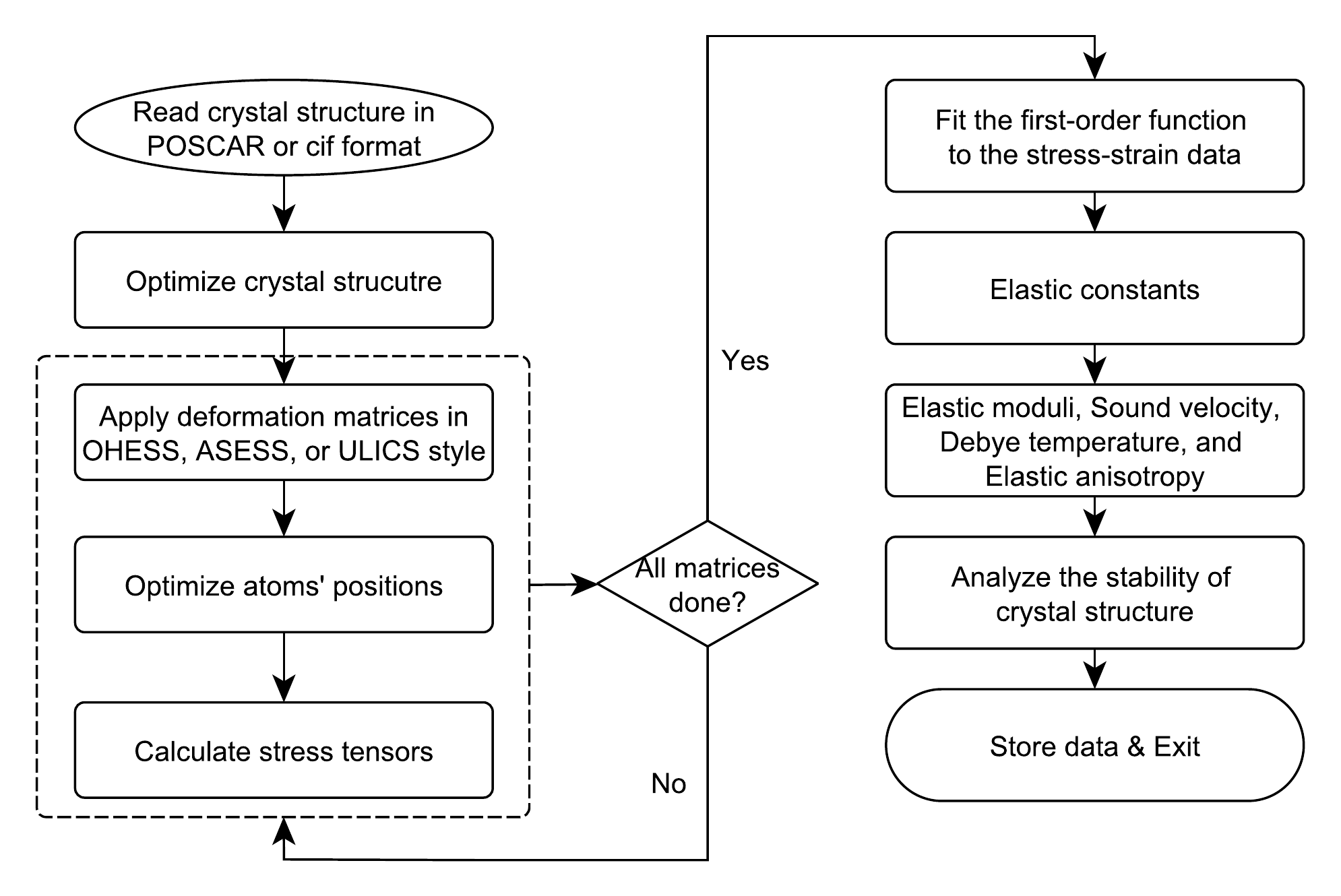}
	\caption{The flowchart of \textsc{ElasTool}}
	\label{flowchart}
\end{figure*}

The flowchart of \textsc{ElasTool} package is illustrated in
Fig.\ref{flowchart}. First, \textsc{ElasTool} reads the crystal structure
file in either POSCAR or cif format, and \tmcolor{pastelyellow}{}calls VASP \cite{Kresse1996,Kresse1999} to
optimize the initial structure at fixed pressure or volume specified in the
INCARs file. For the optimized crystal lattice, \textsc{ElasTool} then
applies different types of deformation matrices according to its symmetry and
optimizes all atoms' positions. Then, all the stress tensors corresponding to
each deformation matrix are calculated and collected. Subsequently,
\textsc{ElasTool} fits the first-order function to the collected
stress-strain data to deduce the full elastic constants. The elastic moduli
are calculated according to Eqs. \ref{bvrh}-\ref{nu}. The Debye temperature is
calculated from Eq. \ref{td}. The sound velocity is calculated from Eq.
\ref{vp}-\ref{vm}. The elastic anisotropy is analyzed by Eqs. \ref{ac} and
\ref{au}. The mechanical stability of this crystal structure is analyzed
according to Born stability criteria proposed in Refs.
{\cite{Wu2007,Mouhat2014}}. Finally, \textsc{ElasTool} stores all the data
calculated to the {\tmsamp{elastool.out}} file and exits.

\subsection{ Installation requirements}

\textsc{ElasTool} is based on Python and its installation is very easy.
But before the installation of \textsc{ElasTool}, the necessary libraries
should be installed first. The following packages are required:

$\bullet$ Python3.5 or later.

$\bullet$ NumPy.

$\bullet$ Spglib.

$\bullet$ ASE.

$\bullet$ Pandas.

$\bullet$ VASP.

Python 3 is the basic language environment for \textsc{ElasTool} running.
\tmverbatim{Numpy} is for numerical calculations. \tmverbatim{Spglib} is for
the automatic determination of symmetry of crystal structure. \tmverbatim{ASE}
is used to manipulate the crystal structures. \tmverbatim{Pandas} is for the
statistics of stress tensors calculated. At present, \textsc{ElasTool}
interfaces to \tmverbatim{VASP} package for calculating the accurate stresses
of strained crystal. But the interfaces to other DFT packages can also be
easily implemented.

\subsection{Installation}

In the Python 3 environment, the necessary libraries can be installed via
{\tmsamp{pip}} command, \textit{e.g.}, {\tmsamp{pip install numpy}}. For the
construction of the environment and libraries, the \tmverbatim{miniconda3}
management platform of Python packages is highly recommended. After installing
\tmverbatim{miniconda3}, the basic Python 3 language environment is
constructed and the other libraries can be installed via either
{\tmsamp{conda}} or {\tmsamp{pip}} commands. For example, one can install
numpy via {\tmsamp{conda install -c conda-forge numpy}}.

\subsection{Run}

To run \textsc{ElasTool}, one only needs to execute {\tmsamp{elastool}} in
the working directory. \textsc{ElasTool} will automatically prepare
necessary files for calculating the stresses of crystal under deformation, and
then call VASP to optimize initial crystal structure and calculate the
stresses for each deformation defined by OHESS, ASESS, or ULICS. Finally,
\textsc{ElasTool} analyzes the stress-strain relationship according to Hooke's
law and calculates all the elastic constants.

\section{Input/output files\label{rd}}\label{iop}

\subsection{Input files}

\textsc{ElasTool} needs one main input file for setting the calculation
details of elastic constants. The main input file is named
{\tmsamp{elatool.in}}. The crystal structure file is provided either in POSCAR
or cif format for reading in the structure information of the crystal. For
VASP stress tensor calculations, INCARs, KPOINTS-static, KPOINTS-dynamic, and
POTCAR-XX files are also necessary. In the INCARs file, several INCAR files of
VASP are collected for optimization, the static calculation, and the molecular
dynamics simulations of high-temperature elastic constants. The
KPOINTS-static file is for the structure optimization and the static
calculation of stress tensors. The KPOINTS-dynamic file is for the calculation
of high-temperature elastic constants using molecular dynamics. XX in the
POTCAR file name is the abbreviated name for an element of the crystal.

\section{Input parameters}\label{ip}

There are totally 10 controlling parameters of \textsc{ElasTool}, as
listed in Table.\ref{parameters}. The {\tmsamp{run\_mode}} sets the running
mode for ElasTool, 1 for automatic run, 2 for pre-processing, and 3 for
post-processing. If {\tmsamp{run\_mode = 2 or 3}}, one should ensure the
structure has already been optimized at fixed pressure or volume,
{\tmem{i.e.}} both the CONTCAR and OUTCAR files are in {\tmsamp{./OPT}}
directory. In running mode 2, \textsc{ElasTool} will directly prepare all
the necessary files for calculating stress tensors. After all the stress
tensors calculations are finished, run mode 3 can analyze the output files and
extract stress tensors, and then fit the first-order function to the
stress-tensor data to obtain elastic constants. The {\tmsamp{dimensional}}
defines the dimensional of the system, 2D or 3D. If the system is 2D,
\textsc{ElasTool} supposes the layered sheet in the {\tmem{xy}}-plane. The
{\tmsamp{structure\_file}} specifies the original crystal structure file in
POSCAR (.vasp) or cif (.cif) format. The {\tmsamp{if\_conventional\_cell}}
determines the usage of primitive cell (no) or conventional cell (yes). The
{\tmsamp{method\_stress\_statistics}} chooses the elastic constants
calculation method, static or dynamic, {\tmsamp{static}} for 0 K elastic
constants, {\tmsamp{dynamic}} for high-temperature. The static method uses the
static stress to compute elastic constants, while the dynamic method deduces
elastic constants from the thermal stresses obtained by molecular dynamics
simulations. The {\tmsamp{strains\_matrix}} defines the type of strain-matrix
set, OHESS, ASESS, or ULICS. The {\tmsamp{strains\_list}} gives one or more
strains for calculating stresses via the strain-matrix set of OHESS, ASESS, or
ULICS. The {\tmsamp{repeat\_num}} controls how to build a supercell from the
primitive or conventional cell defined by {\tmsamp{if\_conventional\_cell}}
for the dynamic method. The {\tmsamp{num\_last\_samples }}is the number of
last MD steps to average thermal stresses. The
{\tmsamp{parallel\_submit\_command}} is the parallel submitting command of
{\tmem{ab initio}} code, e.g. VASP.

\begin{table*}[h]
	\centering{
			\caption{The controlling parameters and possible values of
			\textsc{ElasTool}\label{parameters}}
	\begin{tabular}{ll}
		\hline
		\hline
		Parameters & Values\\
		\hline
		{\tmsamp{run\_mode}} & {\tmsamp{1/2/3}}\\
		{\tmsamp{dimensional}} & {\tmsamp{2D/3D}}\\
		{\tmsamp{structure\_file}} & {\tmsamp{file name ended with .vasp or
				.cif}}\\
		{\tmsamp{if\_conventional\_cell}} & {\tmsamp{yes/no}}\\
		{\tmsamp{method\_stress\_statistics}} & {\tmsamp{static/dynamic}}\\
		{\tmsamp{strains\_matrix}} & {\tmsamp{ohess/asess/ulics}}\\
		{\tmsamp{strains\_list}} & {\tmsamp{one or more numbers}}\\
		{\tmsamp{repeat\_num}} & {\tmsamp{3 integers}}\\
		{\tmsamp{num\_last\_samples}} & {\tmsamp{1 integer}}\\
		{\tmsamp{parallel\_submit\_command}} & {\tmsamp{DFT parallel run
				command}}\\
		\hline
		\hline
	\end{tabular}
}
\end{table*}

\section{Example of run}\label{exa}\label{exa}

The best way to learn \textsc{ElasTool} is to start from the examples.
\textsc{ElasTool} can calculate zero-temperature and high-temperature
elastic constants. The zero-temperature calculations can be conducted by
static stress computation. The high-temperature elastic constants can be
derived by molecular dynamics simulations.

\subsection{Zero-temperature elastic constants}

We take the 0 K elastic constants calculation of diamond as the static
example. The content of the input file {\tmsamp{elastool.in}} is as follows.

\begin{small}
\tmsamp{run\_mode = 1}

\tmsamp{dimensional = 3D}

\tmsamp{structure\_file = diamond.cif}

\tmsamp{if\_conventional\_cell = no}

\tmsamp{method\_stress\_statistics = static}

\tmsamp{strains\_matrix = ohess}

\tmsamp{strains\_list = -0.06 -0.03 0.03 0.06}

\tmsamp{\#repeat\_num = 1 1 1}

\tmsamp{\#num\_last\_samples = 1}

\tmsamp{parallel\_submit\_command = mpirun -np 28 vasp544}
\end{small}

\subsection{High-temperature elastic constants}

The high-temperature elastic constants calculation of metal copper is the
high-temperature example. We build a $3 \times 3 \times 3$ supercell from the
conventional cell of face-centered-cubic of Cu and then perform long-time MD
simulations defined in the INCAR-dynamic file. Because MD is very time
consuming, there is only one strain of -0.06 is used. The last 500 MD steps
are used to average thermal stresses.

\begin{small}
\tmsamp{run\_mode = 1}

\tmsamp{dimensional = 3D}

\tmsamp{structure\_file = CONTCAR.vasp}

\tmsamp{if\_conventional\_cell = yes}

\tmsamp{method\_stress\_statistics = dynamic}

\tmsamp{strains\_matrix = ohess}

\tmsamp{strains\_list = -0.06}

\tmsamp{repeat\_num = 3 3 3}

\tmsamp{num\_last\_samples = 500}

\tmsamp{parallel\_submit\_command = mpirun -np 28 vasp544}
\end{small}

\subsection{Output files}

The {\tmsamp{elastool.out}} file is the unique output file of
\textsc{ElasTool}. It includes the calculated elastic constants data, the
elastic moduli, the sound velocity, the Debye temperature, the elastic
anisotropy, and the stability analysis results of the crystal structure based
on Born elastic criteria. The printed information on the screen of the diamond
example is as follows.

\begin{small}
\tmsamp{Reading controlling parameters from elastool.in...}

\tmsamp{Calculating stresses using the OHESS strain matrices...}

\tmsamp{strain = -0.060}

\tmsamp{strain = -0.030}

\tmsamp{strain = 0.030}

\tmsamp{strain = 0.060}

\tmsamp{Fitting the first-order function to the collected}

\tmsamp{stress-strain data according to Hooke's law...}

\tmsamp{The finnal results are as follows:}

\tmsamp{+==========================================+}

\tmsamp{\textbar This is a 3D Cubic lattice.  \ \ \ \ \ \ \ \ \ \ \ \ \ \ \textbar}

\tmsamp{\textbar -------------------------------------------\textbar}

\tmsamp{\textbar Mean Pressure = -0.08 GPa \ \ \ \ \ \ \ \ \ \ \ \ \ \ \ \ \ \textbar}

\tmsamp{\textbar -------------------------------------------\textbar}

\tmsamp{\textbar Elastic constants: \ \ \ \ \ \ \ \ \ \ \ \ \ \ \ \ \ \ \ \ \ \ \
\textbar}

\tmsamp{\textbar C11 = 1055.04 GPa \ \ \ \ \ \ \ \ \ \ \ \ \ \ \ \ \ \ \ \ \ \ \ \ \ \textbar}

\tmsamp{\textbar C12 = 136.56 GPa \ \ \ \ \ \ \ \ \ \ \ \ \ \ \ \ \ \ \ \ \ \ \ \ \ \
\textbar}

\tmsamp{\textbar C44 = 567.76 GPa \ \ \ \ \ \ \ \ \ \ \ \ \ \ \ \ \ \ \ \ \ \ \ \ \ \
\textbar}

\tmsamp{\textbar -------------------------------------------\textbar}

\tmsamp{\textbar Elastic moduli: \ \ \ \ \ \ \ \ \ \ \ \ \ \ \ \ \ \ \ \ \ \ \ \
\ \ \textbar}

\tmsamp{\textbar B\_V = 442.72 GPa\ \ \ \ \ \ \ \ \ \ \ \ \ \ \ \ \ \ \ \ \ \ \ \ \ \ \ \textbar}

\tmsamp{\textbar B\_R = 442.72 GPa \ \ \ \ \ \ \ \ \ \ \ \ \ \ \ \ \ \ \ \ \ \ \ \ \ \
\textbar}

\tmsamp{\textbar G\_V = 524.36 GPa \ \ \ \ \ \ \ \ \ \ \ \ \ \ \ \ \ \ \ \ \ \ \ \ \ \
\textbar}

\tmsamp{\textbar G\_R = 518.73 GPa \ \ \ \ \ \ \ \ \ \ \ \ \ \ \ \ \ \ \ \ \ \ \ \ \ \
\textbar}

\tmsamp{\textbar B\_VRH = 442.72 GPa \ \ \ \ \ \ \ \ \ \ \ \ \ \ \ \ \ \ \ \ \ \ \ \ \textbar}

\tmsamp{\textbar G\_VRH = 521.54 GPa\ \ \ \ \ \ \ \ \ \ \ \ \ \ \ \ \ \ \ \ \ \ \ \ \ \textbar}

\tmsamp{\textbar Young's modulus (E) = 1123.47 GPa \ \ \ \ \ \ \ \ \ \textbar}

\tmsamp{\textbar Possion's ratio (V) = 0.0771 \ \ \ \ \ \ \ \ \ \ \ \ \ \ \textbar}

\tmsamp{\textbar -------------------------------------------\textbar}

\tmsamp{\textbar Sound velocity:\ \ \ \ \ \ \ \ \ \ \ \ \ \ \ \ \ \ \ \ \ \ \ \ \ \ \ \ \textbar}

\tmsamp{\textbar V\_S = 12.20 Km/s \ \ \ \ \ \ \ \ \ \ \ \ \ \ \ \ \ \ \ \ \ \ \ \ \ \ \textbar}

\tmsamp{\textbar V\_B = 11.24 Km/s \ \ \ \ \ \ \ \ \ \ \ \ \ \ \ \ \ \ \ \ \ \ \ \ \ \ \textbar}

\tmsamp{\textbar V\_P = 18.03 Km/s \ \ \ \ \ \ \ \ \ \ \ \ \ \ \ \ \ \ \ \ \ \ \ \ \ \ \textbar}

\tmsamp{\textbar V\_M = 13.31 Km/s \ \ \ \ \ \ \ \ \ \ \ \ \ \ \ \ \ \ \ \ \ \ \ \ \ \ \textbar}

\tmsamp{\textbar -------------------------------------------\textbar}

\tmsamp{\textbar Debye temperature: \ \ \ \ \ \ \ \ \ \ \ \ \ \ \ \ \ \ \ \ \ \ \
\textbar}

\tmsamp{\textbar T\_D = 1761.03 K \ \ \ \ \ \ \ \ \ \ \ \ \ \ \ \ \ \ \ \ \ \ \ \ \ \ \
\textbar}

\tmsamp{\textbar -------------------------------------------\textbar}

\tmsamp{\textbar Elastic anisotropy: \ \ \ \ \ \ \ \ \ \ \ \ \ \ \ \ \ \ \ \ \ \
\textbar}

\tmsamp{\textbar A\_U = 0.0542 \ \ \ \ \ \ \ \ \ \ \ \ \ \ \ \ \ \ \ \ \ \ \ \ \ \ \ \ \ \
\textbar}

\tmsamp{\textbar A\_C = 0.0054 \ \ \ \ \ \ \ \ \ \ \ \ \ \ \ \ \ \ \ \ \ \ \ \ \ \ \ \ \ \
\textbar}

\tmsamp{\textbar -------------------------------------------\textbar}

\tmsamp{\textbar Structure stability analysis... \ \ \ \ \ \ \ \ \ \ \textbar}

\tmsamp{\textbar This structure is mechanically STABLE. \ \ \ \  \textbar}

\tmsamp{+==========================================+}

\tmsamp{Results are also saved in the elastool.out file.}

\tmsamp{Well done! GOOD LUCK!}
\end{small}

\section{Conclusions}

In summary, we here \label{conclusion}introduced the automatic calculation
toolkit of second-order elastic constants, \textsc{ElasTool}. First, we
recalled the theoretical background and computation method of elastic
constants, elastic moduli, sound velocity, elastic anisotropy, and Debye
temperature. Then, we described the structure of \textsc{ElasTool} package
and its installation. Subsequently, we detailed the necessary input files and
the controlling parameters of \textsc{ElasTool}. Finally, the calculation
examples of 0 K and high-temperature elastic constants are illustrated in
detail. The running output of \textsc{ElasTool} is also presented.
Overall, we developed a useful toolkit for calculating elastic constants, as is
of great significance for either the exploration of materials' elasticity or
high-throughput new materials design.

\section{Acknowledgments}

We acknowledge the support from the National Natural Science Foundation of
China (41574076), the Key Research Scheme of Henan Universities (18A140024),
and the Research Scheme of LYNU Innovative Team under Grant No. B20141679.

%\bibliography{Refs}

\begin{thebibliography}{0}
\bibitem{1} \url{https://numpy.org/} 
\bibitem{2} \url{https://atztogo.github.io/spglib/}
\bibitem{3} \url{https://wiki.fysik.dtu.dk/ase/}
\bibitem{4} \url{https://pandas.pydata.org/}
\bibitem{5} \url{https://www.vasp.at/}
\bibitem{6} G. Kresse, J. Furthmüller, Efficient iterative schemes for ab initio total-energy calculations using a plane-wave basis set, Phys. Rev. B 54 (1996)
	11169.
\bibitem{7} G. Kresse, D. Joubert, From ultrasoft pseudopotentials to the projector
	augmented-wave method, Phys. Rev. B 59 (1999) 1758.
\end{thebibliography}

\begin{thebibliography}{10}
	\expandafter\ifx\csname url\endcsname\relax
	\def\url#1{\texttt{#1}}\fi
	\expandafter\ifx\csname urlprefix\endcsname\relax\def\urlprefix{URL }\fi
	\expandafter\ifx\csname href\endcsname\relax
	\def\href#1#2{#2} \def\path#1{#1}\fi
	
	\bibitem{Ashcroft1976}
	N.~W. Ashcroft, N.~D. Mermin, Solid State Physics, Harcourt College Publishers,
	New York, 1976.
	
	\bibitem{Tehrani2019}
	A.~M. Tehrani, J.~Brgoch, Hard and superhard materials: A computational
	perspective, J. Solid State Chem. 271 (2019) 47.
	\newblock \href {https://doi.org/10.1016/j.jssc.2018.10.048}
	{\path{doi:10.1016/j.jssc.2018.10.048}}.
	
	\bibitem{Lay1995}
	T.~Lay, T.~C. Wallace, Modern Global Seismology, Elsevier Inc., Amsterdam,
	1995.
	
	\bibitem{Tatsumi2003}
	K.~Tatsumi, I.~Tanaka, K.~Tanaka, H.~Inui, M.~Yamaguchi, H.~Adachi, M.~Mizuno,
	Elastic constants and chemical bonding of {LaNi}$_{5}$ and
	{LaNi}$_{5}$h$_{7}$ by first principles calculations, J. Phys.: Condens.
	Matt. 15 (2003) 6549.
	\newblock \href {https://doi.org/10.1088/0953-8984/15/38/021}
	{\path{doi:10.1088/0953-8984/15/38/021}}.
	
	\bibitem{Liu2014}
	Z.~L. Liu, Muse: Multi-algorithm collaborative crystal structure prediction,
	Comput. Phys. Commun. 185 (2014) 1893.
	\newblock \href {https://doi.org/10.1016/j.cpc.2014.03.017}
	{\path{doi:10.1016/j.cpc.2014.03.017}}.
	
	\bibitem{Wang2010}
	Y.~Wang, J.~Lv, L.~Zhu, Y.~Ma, Crystal structure prediction via particle-swarm
	optimization, Phys. Rev. B 82 (2010) 094116.
	\newblock \href {https://doi.org/10.1103/PhysRevB.82.094116}
	{\path{doi:10.1103/PhysRevB.82.094116}}.
	
	\bibitem{Wang2012}
	Y.~Wang, J.~Lv, L.~Zhu, Y.~Ma, Calypso: A method for crystal structure
	prediction, Comput. Phys. Commun. 183 (2012) 2063.
	\newblock \href {https://doi.org/10.1016/j.cpc.2012.05.008}
	{\path{doi:10.1016/j.cpc.2012.05.008}}.
	
	\bibitem{Glass2006}
	C.~W. Glass, A.~R.Oganov, N.~Hansen, Uspex—evolutionary crystal structure
	prediction, Comput. Phys. Commun. 175.
	
	\bibitem{Lonie2011}
	D.~C. Lonie, E.~Zurek, Xtalopt: An open-source evolutionary algorithm for
	crystal structure prediction, Comput. Phys. Commun. 182.
	
	\bibitem{Pickard2011}
	C.~J. Pickard, R.~J. Needs, Ab initio random structure searching, J. Phys.:
	Condens. Matt. 23.
	
	\bibitem{Wu2007}
	Z.~J. Wu, E.~J. Zhao, H.~P. Xiang, X.~F. Hao, X.~J. Liu, J.~Meng, Crystal
	structures and elastic properties of superhard $\mathrm{Ir}{\mathrm{n}}_{2}$
	and $\mathrm{Ir}{\mathrm{n}}_{3}$ from first principles, Phys. Rev. B 76
	(2007) 054115.
	\newblock \href {https://doi.org/10.1103/PhysRevB.76.054115}
	{\path{doi:10.1103/PhysRevB.76.054115}}.
	
	\bibitem{Mouhat2014}
	F.~Mouhat, F.~X. Coudert, Necessary and sufficient elastic stability conditions
	in various crystal systems, Phys. Rev. B 90 (2014) 224104.
	\newblock \href {https://doi.org/10.1103/PhysRevB.90.224104}
	{\path{doi:10.1103/PhysRevB.90.224104}}.
	
	\bibitem{Nye1985}
	J.~F. Nye, Physical Properties of Crystals, Oxford University Press, Oxford,
	1985.
	
	\bibitem{Born1954}
	M.~Born, K.~Huang, Dynamical Theory of Crystal Lattices, Oxford University
	Press, Oxford, 1954.
	
	\bibitem{Liu2017}
	Z.~L. Liu, H.~Jia, R.~Li, X.~L. Zhang, L.~C. Cai, Unexpected coordination
	number and phase diagram of niobium diselenide under compression, Phys. Chem.
	Chem. Phys. 19 (2017) 13219.
	\newblock \href {https://doi.org/10.1039/c7cp00805h}
	{\path{doi:10.1039/c7cp00805h}}.
	
	\bibitem{Yu2010}
	R.~Yu, J.~Zhu, H.~Q. Ye, Calculations of single-crystal elastic constants made
	simple, Comput. Phys. Commun. 181 (2010) 671.
	\newblock \href {https://doi.org/10.1016/j.cpc.2009.11.017}
	{\path{doi:10.1016/j.cpc.2009.11.017}}.
	
	\bibitem{Jong2015}
	M.~D. Jong, W.~Chen, T.~Angsten, A.~Jain, R.~Notestine, A.~Gamst, M.~Sluiter,
	C.~K. Ande, S.~van~der Zwaag, J.~J. Plata, C.~Toher, S.~Curtarolo, G.~Ceder,
	K.~A. Persson, M.~Asta, Charting the complete elastic properties of inorganic
	crystalline compounds, Scientific Data 2 (2015) 150009.
	\newblock \href {https://doi.org/10.1038/sdata.2015.9}
	{\path{doi:10.1038/sdata.2015.9}}.
	
	\bibitem{Choudhary2018}
	K.~Choudhary, G.~Cheon, E.~Reed, F.~Tavazza, Elastic properties of bulk and
	low-dimensional materials using van der waals density functional, Phys. Rev.
	B 98.
	
	\bibitem{Hohenberg1964}
	P.~Hohenberg, W.~Kohn, Inhomogeneous electron gas, Phys. Rev. 136 (1964) B864.
	\newblock \href {https://doi.org/10.1103/PhysRev.136.B864}
	{\path{doi:10.1103/PhysRev.136.B864}}.
	
	\bibitem{Kohn1965}
	W.~Kohn, L.~J. Sham, Self-consistent equations including exchange and
	correlation effects, Phys. Rev. 140 (1965) A1133.
	\newblock \href {https://doi.org/10.1103/PhysRev.140.A1133}
	{\path{doi:10.1103/PhysRev.140.A1133}}.
	
	\bibitem{Wang2019}
	V.~Wang, N.~Xu, J.~C. Liu, G.~Tang, W.~T. Geng, Vaspkit: A pre- and
	post-processing program for vasp code (arxiv:1908.08269v2) (2019).
	\newblock \href {http://arxiv.org/abs/arXiv:1908.08269}
	{\path{arXiv:arXiv:1908.08269}}.
	
	\bibitem{Golesorkhtabar2013}
	R.~Golesorkhtabar, P.~Pavone, J.~Spitaler, P.~Puschnig, C.~Draxl, Elastic: A
	tool for calculating second-order elastic constants from first principles,
	Comput. Phys. Commun. 184 (2013) 1861.
	\newblock \href {https://doi.org/10.1016/j.cpc.2013.03.010}
	{\path{doi:10.1016/j.cpc.2013.03.010}}.
	
	\bibitem{Perger2009}
	W.~F. Perger, J.~Criswell, B.~Civalleri, R.~Dovesi, Ab-initio calculation of
	elastic constants of crystalline systems with the crystal code, Comput. Phys.
	Commun. 180 (2009) 1753.
	\newblock \href {https://doi.org/10.1016/j.cpc.2009.04.022}
	{\path{doi:10.1016/j.cpc.2009.04.022}}.
	
	\bibitem{Sinko2002}
	G.~V. Sin’ko, N.~A. Smirnov, Ab initio calculations of elastic constants and
	thermodynamic properties of bcc, fcc, and hcp al crystals under pressure, J.
	Phys.: Condens. Matter 14 (2002) 6989.
	\newblock \href {https://doi.org/10.1088/0953-8984/14/29/301}
	{\path{doi:10.1088/0953-8984/14/29/301}}.
	
	\bibitem{Liu2020}
	Z.~L. Liu, High-efficiency calculation of elastic constants enhanced by the
	optimized strain-matrix sets(arxiv:2002.00005) (2020).
	\newblock \href {http://arxiv.org/abs/arXiv:2002.00005}
	{\path{arXiv:arXiv:2002.00005}}.
	
	\bibitem{Hill1952}
	R.~Hill, The elastic behaviour of a crystalline aggregate, Proc. Phys. Soc.
	London , 350 65.
	
	\bibitem{Chung1967}
	D.~H. Chung, W.~R. Buessem, The elastic anisotropy of crystals, J. of Appl.
	Phys. 38.
	
	\bibitem{Ranganathan2008}
	S.~I. Ranganathan, M.~Ostoja-Starzewski, Universal elastic anisotropy index,
	Phys. Rev. Lett. 101.
	
	\bibitem{Kresse1996}
	G.~Kresse, J.~Furthm\"uller, Efficient iterative schemes for ab initio
	total-energy calculations using a plane-wave basis set, Phys. Rev. B 54
	(1996) 11169.
	\newblock \href {https://doi.org/10.1103/PhysRevB.54.11169}
	{\path{doi:10.1103/PhysRevB.54.11169}}.
	
	\bibitem{Kresse1999}
	G.~Kresse, D.~Joubert, From ultrasoft pseudopotentials to the projector
	augmented-wave method, Phys. Rev. B 59 (1999) 1758.
	\newblock \href {https://doi.org/10.1103/PhysRevB.59.1758}
	{\path{doi:10.1103/PhysRevB.59.1758}}.
	
\end{thebibliography}
\bibliographystyle{elsarticle-num}

\end{document}